\documentclass[preprint,12pt]{elsarticle}

\usepackage{amsmath}
\usepackage{amsfonts}
\usepackage{amssymb}
\usepackage{physics}
\usepackage{hyperref}
\usepackage{braket}
\usepackage{pdfpages}
\usepackage{transparent}
\usepackage{float}
\usepackage{booktabs}
\usepackage{threeparttable}
\usepackage[figuresright]{rotating}
\usepackage{extarrows}
\usepackage{mathrsfs}

\hypersetup{hidelinks}

\journal{Fundamental Research on Jan 17, 2025}

\begin{document}

\newcommand{\ii}{\mathrm{i}}
\newcommand{\ee}{\mathrm{e}}

\begin{frontmatter}



\title{Constraint Phase Space Formulations for Finite-State Quantum Systems: The Relation between Commutator Variables and Complex Stiefel Manifolds\tnoteref{label1}} 







\author[inst1]{Youhao Shang\fnref{fn1}}

\author[inst1]{Xiangsong Cheng\fnref{fn1}}

\author[inst1]{Jian Liu\corref{cor1}}
\ead{jianliupku@pku.edu.cn}

\cortext[cor1]{Corresponding author}

\fntext[fn1]{Both authors contributed equally.}

\affiliation[inst1]{
    organization={Beijing National Laboratory for Molecular Sciences, Institute of Theoretical and Computational Chemistry, College of Chemistry and Molecular Engineering, Peking University},
    city={Beijing},
    postcode={100871},
    country={China}
}

\begin{abstract}
    {We have recently developed the \textit{constraint} coordinate-momentum \textit{phase space} (CPS) formulation for finite-state quantum systems. It has been implemented for the electronic subsystem in nonadiabatic transition dynamics to develop practical trajectory-based approaches. In the generalized CPS formulation for the mapping Hamiltonian of the classical mapping model with commutator variables (CMMcv) method [\textit{J. Phys. Chem. A} \textbf{2021}, 125, 6845-6863], each {connected} component of the generalized CPS is the \textit{complex Stiefel manifold} labeled by the eigenvalue set of the mapping kernel. Such a phase space structure allows for exact trajectory-based dynamics for pure discrete (electronic) degrees of freedom (DOFs), where the equations of motion of each trajectory are isomorphic to the time-dependent Schr\"odinger equation.  We employ covariant kernels {within the generalized CPS framework} to develop two approaches that naturally yield exact evaluation of time correlation functions (TCFs) for pure discrete (electronic) DOFs.  In addition, we briefly discuss the phase space mapping formalisms where the contribution of each trajectory to the integral expression of the {TCF} of population dynamics is strictly positive semi-definite.  The generalized CPS formulation also indicates that the equations of motion in phase space mapping model I of our previous work [\textit{J. Chem. Phys.} \textbf{2016}, 145, 204105; \textbf{2017}, 146, 024110; \textbf{2019}, 151, 024105] lead to a complex Stiefel manifold $\mathrm{U}(F)/\mathrm{U}(F-2)$.
    It is expected that the generalized CPS formulation has implications for simulations of both nonadiabatic transition dynamics and many-body quantum dynamics for spins/bosons/fermions.}
\end{abstract}




\end{frontmatter}


\section{Introduction}

Phase space with coordinate-momentum variables bridges a rigorous formulation of classical mechanics\cite{Arnold1989, Nolte2010} and an exact interpretation of quantum mechanics\cite{Lee1995, Polkovnikov2010, Liu2011c, Liu2011b, Liu2011e, He2022}.  Since the pioneering work of Weyl and Wigner for introducing the one-to-one correspondence mapping between quantum operators and phase space functions\cite{Weyl1927, Wigner1932, Groenewold1946, Moyal1949, Husimi1940, Cohen1966}, the coordinate-momentum phase space formulation has been well established as an exact interpretation of quantum mechanics for systems with continuous degrees of freedom (DOFs).  Similar to phase space (with coordinate-momentum variables) of classical mechanics, quantum phase space for continuous DOFs involves infinite volume.  Such a relation makes it possible to employ the Wigner or Husimi phase space \cite{Wigner1932, Husimi1940} to develop a few trajectory-based practical approaches for describing nuclear quantum dynamics effects of complex molecular systems. E.g., the linearized semiclassical initial value representation (LSC-IVR)/classical Wigner approach\cite{Wang1998, Sun1998, Liu2007, Liu2009a, Shi2003, Pollak1998, Shao1998, Hernandez1998, Poulsen2003, Liu2015}, forward-backward semiclassical dynamics\cite{Shao1999, Shao1999a}, path integral Liouville dynamics and other phase space quantum dynamics approaches\cite{Liu2011c, Liu2011b, Liu2011e, Liu2014, Liu2016b, Liu2016c, Zhang2020, Liu2021} based on the (general) Moyal bracket expression of quantum Liouville theorem\cite{Groenewold1946, Moyal1949}.  

Because a typical (large) molecular system often involves (coupled) rotational, vibrational, and hindered translational motion, it is more practical to use the continuous coordinate space rather than the Hilbert space with dense states to deal with nuclear DOFs.  In comparison, since the energy gap between ground and excited electronic states of our interest is usually much larger, electronic DOFs are often described by the discrete adiabatic or (quasi-)diabatic state representation instead of continuous electronic coordinate space.  When the nonadiabatic transition process includes coupled electronic and nuclear motion, it presents a composite system, where nuclear DOFs are depicted by continuous variables while electronic DOFs are described by discrete (electronic-state) variables\cite{He2022}.  Such a composite system often goes beyond the scope of conventional coordinate-momentum phase space formulations\cite{Weyl1927, Wigner1932, Groenewold1946, Moyal1949, Husimi1940, Cohen1966, Liu2011c}, because a finite-state quantum system has no true classical analogy\cite{Sakurai2020}.

Nevertheless, diversified isomorphisms or analogies have been proposed in ever-increasing levels of abstraction. A few heuristic quantum-classical analogies\cite{Miller1978, McCurdy1979, Meyer1979, Meyer1979a, Meyer1980, Kryvohuz2005, Miller2016a} have been proposed. E.g., the analogy of classical angular momentum to spin\cite{Meyer1979a, Cotton2015} and to the many-electron Hamiltonian\cite{Li2012}, the Bohr-Sommerfeld quantization with action-angle variables for Wigner functions\cite{Kryvohuz2005, Miller2016a}, and the ``classical electron-analog" in ref. \cite{Meyer1979} that yielded the celebrated Meyer-Miller mapping Hamiltonian for nonadiabatic dynamics \cite{Tang2019, Hu2021a, Lin2022, Xie2018, Gao2020b}. 

Rigorous isomorphisms of pure finite-state systems fall into two main categories. The first category employs the Schwinger oscillator theory of angular momentum\cite{Schwinger1965, Stock1997} or the Holstein-Primakoff transformation\cite{Holstein1940, Garbaczewski1978, Blaizot1978a}. For example, Stock and Thoss have used the Schwinger oscillator theory of angular momentum\cite{Schwinger1965} to prove the Meyer-Miller mapping Hamiltonian model\cite{Meyer1979} is exact in quantum mechanics\cite{Stock1997}, so it is also often called the Meyer-Miller-Stock-Thoss (MMST) model. The drawback of this category of theory is that only a physical subspace can be involved in the mapping{\cite{Thoss1999,Liu2016,Liu2017}}, which is exact in quantum mechanics but meet problems when (trajectory-based) approximations are introduced such that the larger or full physical space is effectively explored for composite/nonadiabatic systems{ \cite{Liu2016, Liu2017, Liu2021, He2022}}.  E.g., the LSC-IVR for both electronic and nuclear DOFs of the MMST mapping Hamiltonian model\cite{Sun1998} leads to poor performance \cite{Coronado2001, Ananth2007}. The Holstein-Primakoff transformation\cite{Holstein1940, Garbaczewski1978, Blaizot1978a} and the Dyson-Maleev transformation\cite{Dyson1956, Dyson1956a, Maleev1957} meet similar problems for composite/nonadiabatic systems. In addition, the Jordan-Wigner transformation\cite{Jordan1928, Bravyi2002, Seeley2012} constructs the exact mapping between a fermion system and a spin system (which can be isomorphic to a pure finite-state system). It is not clear whether the Jordan-Wigner transformation leads to practically useful quantum dynamics approaches for composite/nonadiabatic systems. The second category consists of exact phase space formulations of finite-state systems. Early contributions to phase space formulations of finite-state systems include the seminar works of Stratonovich in 1956 \cite{Stratonovich1956}, and of Feynman \cite{Feynman1987} and Wootters \cite{Wootters1987} separately in 1987. Subsequent developments for general discrete $F$-state quantum systems in this field follows two main streams: the continuous Stratonovich phase space based on the $\mathrm{SU}(2)$ \cite{Varilly1989,Amiet1991,Amiet2000,Dowling1994,Klimov2002,Klimov2008,Klimov2017} or $\mathrm{SU}(F)$ \cite{Brif1999,Tilma2012,Tilma2016} structure, and the discrete phase space based on the strategy similar to the discrete Fourier transformation \cite{Wootters2006,Cohendet1988,Leonhardt1995,Leonhardt1996,Ruzzi2005,Chaturvedi2006,Gross2006}. Because the $\mathrm{SU}(2)$ or $\mathrm{SU}(F)$ Stratonovich phase space (often) only involves two or $2F-2$ angle variables, the exact equations of motion (EOMs) are highly nonlinear and tedious, where singularities are inevitable and have to be excluded in dynamics, especially when the number of states $F$ is large. More recently, the \textit{constraint} coordinate-momentum \textit{phase space} (CPS) formulation has been introduced for exact mapping of finite-state systems{\cite{Liu2016,Liu2017,He2019,He2021a,He2021,He2022,Liu2021}}. In the CPS formulation, the finite $F$-state quantum system is described by a single CPS\cite{He2021a} or a weighted average over two or more CPSs \cite{He2022}. Embedded in the $(2F)$-dimensional Euclidean space $\mathbb{R}^{2F}$ of the coordinate-momentum variables $\{\mathbf{x},\mathbf{p}\}=\{x_1,\cdots,x_F, p_1,\cdots , p_F\}$, the CPS, which is a $(2F-1)$-sphere, is determined by the following { constraint} relation
\begin{equation}
    \sum_{n=1}^{F}\frac{x_n^2+p_n^2}{2}=1+F\gamma,  \label{eq:CPS-constraint}
\end{equation}
where the value of the phase space parameter $\gamma$ ranges from $-1/F$ to $+\infty$ {\cite{He2019,He2021a}}. In the CPS formulation{\cite{Liu2016,Liu2017,He2019, He2021a}}, the Hamiltonian operator is mapped to a phase space function reminiscent of the Meyer-Miller mapping Hamiltonian \cite{Meyer1979}. In action-angle variables \cite{Cotton2019a}, which are defined as
\begin{equation}
e_n=(x_n^2+p_n^2)/2, \quad \theta_n=\arctan{(p_n/x_n)},\\
\end{equation}
the CPS constraint relation defined in Eq. (\ref{eq:CPS-constraint}) also reads $\sum_{n=1}^{F}e_n=1+F\gamma$, so that the space of action variables is defined via
\begin{equation}
\left\{e_n \geq 0~\forall~ n\in\{1,\cdots,F\}, \quad \sum_{n=1}^{F}e_n=1+F\gamma\right\},\label{eq:simplex}
\end{equation}
which is a simplex composed of all action variables {\cite{He2019,He2021a}}.

Inspired by {refs. \cite{Liu2016,He2019}}, the commutator matrix $\mathbf{\Gamma}$ (which can also be represented by auxiliary coordinate-momentum variables, i.e., commutator variables) is later introduced to a more comprehensive mapping Hamiltonian in ref. \cite{He2021}. The commutator matrix, $\mathbf{\Gamma}$, indicates a more general phase space structure for finite-state systems beyond those described by $\{\mathbf{x},\mathbf{p}\}$ {of Eq. (\ref{eq:CPS-constraint})}.  In this paper, we show that the generalized CPS (with commutator matrix $\mathbf{\Gamma}$) is related to the \textit{complex Stiefel manifold} $V_r(\mathbb{C}^F)\equiv \mathrm{U}(F)/\mathrm{U}(F-r)$ \cite{Atiyah1960,Hatcher2002}, where {Model 2 and Model 1} of refs. {\cite{Liu2016,Liu2017,He2019}} in principle lead to the $r=1$ and $r=2$ cases, respectively.  (Please see more discussion in Appendix B.)  A symplectic structure defines the evolution of trajectories on the generalized CPS.  The exact EOMs are linear on the generalized CPS and involve no singularities.

As pointed out in ref. \cite{He2022}, the three key elements of a trajectory-based quantum dynamics method are
\begin{enumerate}
    \item the EOMs governing the trajectory,
    \item the initial condition for the trajectory, and
    \item the integral expression (of the time correlation function) for the expectation value or ensemble average of the relevant (time-dependent) physical property.
\end{enumerate}
Because the {exact} EOMs on phase space are already defined for pure finite-state systems, we focus on integral expressions for (time-dependent) physical properties. By employing covariant \cite{Varilly1989,Varilly1989a} kernels in the integral expression, we demonstrate two approaches that naturally achieve exact {dynamics results} for pure finite-state systems. One approach involves the covariant estimator for the physical property, and the other employs the covariant kernel for the initial { distribution}. In addition to the two approaches, we show other integral expressions without covariant kernels that produce exact {phase space} dynamics for pure finite-state systems {when $F=2$}\cite{Cotton2016,Cotton2019,Mannouch2023,Cheng2024}.

The paper is organized as follows. Section \ref{sec:general-aspects} provides a concise review of general aspects of the phase space representation of quantum mechanics with mapping formalisms on the CPS. Section \ref{sec:phase-space-structure} presents the complex Stiefel manifold structure of the {generalized CPS} constructed through the covariance condition of the mapping kernel. The exact EOMs of trajectories derived from the symplectic structure is elucidated. Section \ref{sec:correlation-function} shows the two approaches that yield exact {time-dependent properties} of finite-state systems. With the flexibility in choosing the mapping kernels, many existing methods can be unified in the framework of the generalized CPS. We also discuss the integral expressions with non-covariant kernels, where each trajectory makes non-negative contribution to {(electronic)} population dynamics.  Finally, conclusion remarks are presented in Section \ref{sec:final}. Throughout the paper, we use electronically nonadiabatic systems as examples of composite sytems, e.g., electronic DOFs are equivalent to discrete DOFs, and nuclear DOFs represent continuous DOFs.  But we emphasize that it is trivial to generalize the discussion to vibrationally nonadiabatic systems where high-frequency vibrational modes are represented by discrete (state) DOFs while low-frequency modes are depicted by continuous DOFs, or other composite systems in physics, chemistry, biology, materials, {etc.} \cite{HammesSchiffer1994,Kananenka2018,Topaler1994}.  

\section{General Aspects of Phase Space Representation}\label{sec:general-aspects}

The cornerstone of the one-to-one correspondence mapping is the \textit{phase space}, a continuous geometric entity composed of phase points (denoted by $\mathbf{X}$) with an appropriate volume measure (denoted as $\dd\mu(\mathbf{X})$). Throughout this paper, a typical phase space is denoted by $M_{\mathrm{PS}}$ {(a manifold named the phase space)}. Once the phase space has been chosen, quantum operators are {(linearly)} mapped to functions on the phase space. For the (weighted) CPS representation of a discrete $F$-state quantum system \cite{He2019,He2021a,He2021,He2022,Liu2021}, the phase space consists of multiple $(2F-1)$-spheres {$S(\mathbf{x},\mathbf{p};\gamma)$ with radii $\sqrt{2(1+F\gamma)}$ embedded in a $2F$-dimensional Euclidean space}. A point on the phase space is labeled with ``coordinate-momentum" variables
\begin{equation}
    \mathbf{X}=(\mathbf{x},\mathbf{p})=(x_1,\cdots,x_F,p_1,\cdots,p_F).
\end{equation}
The constraint $\mathcal{S}(\mathbf{x},\mathbf{p};\gamma)$ for a sphere $S(\mathbf{x},\mathbf{p};\gamma)$ reads
\begin{equation}
    {\mathcal{S}(\mathbf{x},\mathbf{p};\gamma)=\delta\left(\sum_{n=1}^{F}\frac{x_n^2+p_n^2}{2}-(1+F\gamma)\right),}\label{eq:constraint-eCMM}
\end{equation}
so that the phase points on the sphere satisfy the constraint, Eq. (\ref{eq:CPS-constraint}).
The parameter $\gamma$ labeling the spheres can take values from $-1/F$ to $+\infty$, not limited to the three values $0, 1, (\sqrt{1+F}-1)/F:=\gamma_{\mathrm{w}}$ which are { related to} the P, Q, W versions { of the $\mathrm{SU}(F)$ Stratonovich phase space formulation} \cite{He2021a, He2021}. The integral on CPS is defined as
\begin{equation}
    {\int_{\mathcal{S}(\mathbf{x},\mathbf{p};\gamma)}\mathrm{d}\mu(\mathbf{x},\mathbf{p})g(\mathbf{x},\mathbf{p})=\frac{1}{\Omega(\gamma)}\int F\mathrm{d}\mathbf{x}\mathrm{d}\mathbf{p}\mathcal{S}(\mathbf{x},\mathbf{p};\gamma)g(\mathbf{x},\mathbf{p})}\label{eq:int-eCMM}
\end{equation}
where
\begin{equation}
    {\Omega(\gamma)=\int \mathrm{d}\mathbf{x}\mathrm{d}\mathbf{p}\mathcal{S}(\mathbf{x},\mathbf{p};\gamma)}
\end{equation}
and $\mathrm{d}\mu(\mathbf{x},\mathbf{p})$ is the \textit{invariant measure}. The integral on weighted CPS \cite{He2022} is defined by
\begin{equation}
    {\int_{-1/F}^{+\infty}\dd\gamma\,w(\gamma)\int_{\mathcal{S}(\mathbf{x},\mathbf{p};\gamma)}\mathrm{d}\mu(\mathbf{x},\mathbf{p})g(\mathbf{x},\mathbf{p})},
\end{equation}
where the weight function $w(\gamma)$ represents a \textit{quasi}-probability distribution on the $\gamma$ axis (over different spheres), which obeys
\begin{equation}
    {\int_{-1/F}^{+\infty}\dd\gamma\,w(\gamma)=1}\label{eq:normalization-weight-wMM}
\end{equation}
but can take negative values {of $w(\gamma)$.}

Similar to classical mechanics, we can define \textit{phase trajectories} \cite{Landau1980} stemming from phase points $\mathbf{X}$ on phase space $M_{\mathrm{PS}}$ for finite-state quantum systems. For example, for the finite $F$-state quantum system with Hamiltonian operator
\begin{equation}
    \hat{H} = \sum_{m,n=1}^F H_{nm} \ketbra{n}{m},
\end{equation}
a phase trajectory on the CPS is generated by the Hamilton's EOMs with the mapping Hamiltonian\footnote{Although the EOMs are written in the $2F$-dimensional Euclidean space, the phase space constraint is always satisfied along a phase trajectory starting from a phase point on the CPS $S(\mathbf{x},\mathbf{p};\gamma)$.} {\cite{Dirac1927,Meyer1979,Liu2016,Liu2017}}. The Hamilton's EOMs are also isomorphic to the time-dependent Schr\"odinger equation \cite{Liu2016},
\begin{equation}
    \begin{split}
        &{
        \dot{x}_n = \frac{\partial H_C(\mathbf{x},\mathbf{p};\gamma)}{\partial p_n} = \sum_{m=1}^F H_{nm}p_m,}\\
        &{
        \dot{p}_n = -\frac{\partial H_C(\mathbf{x},\mathbf{p};\gamma)}{\partial x_n} = -\sum_{m=1}^F H_{nm}x_m,}\\
    \end{split} \label{eq:EOM}
\end{equation}
where the mapping Hamiltonian on each component sphere $S(\mathbf{x},\mathbf{p};\gamma)$ is linearly mapped from the {quantum} Hamiltonian $\hat{H}$ of the $F$-state system,
\begin{equation}
    \begin{split}
    H_C(\mathbf{x},\mathbf{p};\gamma) = &\Tr\left[\hat{H}\hat{K}(\mathbf{x},\mathbf{p};\gamma)\right]\\
    = &\sum_{m,n=1}^{F} \left[\frac{1}{2} \left(x_n+\ii p_n\right)\left(x_m-\ii p_m\right) - \gamma \delta_{mn} \right] H_{mn},\\
    \end{split} \label{eq:mapping-Hamiltonian-CMM}
\end{equation}
with the \textit{mapping kernel} {(phase space kernel)} $\hat{K}(\mathbf{x},\mathbf{p};\gamma)$ defined as \cite{He2021a,He2021,Liu2021}
\begin{equation}
    \hat{K}(\mathbf{x},\mathbf{p};\gamma)=\sum_{m,n=1}^{F} \ketbra{n}{m}\left[\frac{1}{2} \left(x_n+\ii p_n\right)\left(x_m-\ii p_m\right) - \gamma \delta_{mn} \right].\label{eq:mapping-kernel-CMM}
\end{equation}
The mapping kernel is normalized such that {the sum over electronic population is 1}.

In the phase space formulation, apart from the mapping relation for the Hamiltonian $H_C(\mathbf{x},\mathbf{p};\gamma)=\Tr\left[\hat{H}\hat{K}(\mathbf{x},\mathbf{p};\gamma)\right]$, {the most important element is} the phase space {integral} expression of the trace of the {two-operator product} $\mathrm{Tr}\left[\hat{\rho}\hat{A}\right]$. We define
\begin{equation}
    \begin{split}
        \rho_C(\mathbf{X}) &= \Tr\left[\hat{\rho}\hat{K}_\rho(\mathbf{X})\right], \\
        \tilde{A}_C(\mathbf{X}) &= \Tr\left[\hat{A}\hat{K}_A(\mathbf{X})\right].
    \end{split}\label{eq:density-kernel-and-observable-kernel}
\end{equation}
Here, $\hat{\rho}$ is the density operator. The operator-valued functions $\hat{K}_\rho(\mathbf{X})$ and $\hat{K}_A(\mathbf{X})$ are denoted as the \textit{density kernel} and the \textit{observable kernel}, respectively. We express the expectation value of a physical observable $\hat{A}$ as the phase space integral\footnote{ As in the CPS formulations in CMM and wMM, the symbol $\mathcal{M}_{\mathrm{PS}}$ denotes for the constraints determining the phase space $M_{\mathrm{PS}}$.},
\begin{equation}
        \langle \hat{A} \rangle_{\hat{\rho}} = \Tr\left[\hat{\rho}\hat{A}\right]= \int_{\mathcal{M}_{\mathrm{PS}}} \dd\mu(\mathbf{X}) \rho_C(\mathbf{X}) \tilde{A}_C(\mathbf{X}).\label{eq:TI-expectation-value}
\end{equation}
Equation (\ref{eq:TI-expectation-value}) is also termed as the \textit{exact mapping condition} {\cite{He2019,He2021}}. For arbitrary $n$, $m$, $k$ and $l$, let $\hat{\rho}=\ketbra{n}{m}$ and $\hat{A}=\ketbra{k}{l}$, the exact mapping condition is equivalent to
\begin{equation}
    {
    \delta_{mk}\delta_{nl}}{=\int_{\mathcal{M}_{\mathrm{PS}}} \dd\mu(\mathbf{X}) \left[\hat{K}_{\rho}(\mathbf{X})\right]_{mn}\left[\hat{K}_{A}(\mathbf{X})\right]_{lk}.}\label{eq:exact-mapping}
\end{equation}

The choice of the kernels $\hat{K}_{\rho}(\mathbf{X})$ and $\hat{K}_{A}(\mathbf{X})$ satisfying Eq. (\ref{eq:exact-mapping}) is \textit{ not} unique \cite{Shang2022thesis}. Based on the CPS, the classical mapping model (CMM) method \cite{He2021} and the weighted mapping model (wMM) method \cite{He2022} define the density kernel $\hat{K}_\rho(\mathbf{x},\mathbf{p};\gamma) =\hat{K}(\mathbf{x},\mathbf{p};\gamma)$, and denote $\hat{K}_A(\mathbf{x},\mathbf{p};\gamma)$ as the inverse mapping kernel $\hat{K}^{-1}(\mathbf{x},\mathbf{p};\gamma)$. For CMM, the observable kernel is given by \cite{He2021a,Liu2021}
\begin{equation}
    \begin{split}
    &\hat{K}_A(\mathbf{x},\mathbf{p};\gamma) = \hat{K}^{-1}(\mathbf{x},\mathbf{p};\gamma)\\
    =&\sum_{m,n=1}^{F} \ketbra{n}{m}\left[\frac{1+F}{2(1+F\gamma)^2} \left(x_n+\ii p_n\right)\left(x_m-\ii p_m\right) - \frac{1-\gamma}{1+F\gamma} \delta_{mn} \right].\\ 
    \end{split} \label{eq:inverse-mapping-kernel-CMM}
\end{equation}
{As a special case of wMM in ref. \cite{He2022}}, when the observable kernel is set to be the same as the density kernel (also the mapping kernel), {such a} choice requires {that} the weight function $w(\gamma)$ {should} satisfy the following condition \cite{He2022}:
\begin{equation}
    \int_{-1/F}^{+\infty}\dd\gamma\,w(\gamma)(F\gamma^2+2\gamma) = 1. \label{eq:weight-function-wMM}
\end{equation}

By involving the time evolution of the phase trajectories, the time correlation function {(TCF)} $\mathrm{Tr}\left[\hat{\rho}\hat{A}(t)\right]$ can also be mapped to an integral on the phase space,
\begin{equation}
    {\mathrm{Tr}\left[\hat{\rho}\hat{A}(t)\right] \equiv \mathrm{Tr}\left[\hat{\rho}\hat{U}^\dagger(t)\hat{A}\hat{U}(t)\right] \mapsto \int_{\mathcal{M}_{\mathrm{PS}}} \dd\mu(\mathbf{X}_0) \rho_C(\mathbf{X}_0) \tilde{A}_C(\mathbf{X}_t),}\label{eq:TD-expectation-value}
\end{equation}
where $\hat{U}(t)$ is the time evolution operator. When the Hamiltonian is time-independent, $\hat{U}(t)=e^{-\ii \hat{H}t}$. The property that the trajectory-based dynamics for pure discrete $F$-state quantum systems is exact for any Hamiltonian is referred to as the \textit{frozen nuclei limit} \cite{Liu2016,He2019} {for nonadiabatic/composite systems} (i.e., continuous DOFs are fixed) in this paper. When the motion of nuclei is frozen, the composite system becomes the pure {(electronic)} $F$-state system. For discrete $F$-state quantum systems, the phase space representation based on $\mathrm{SU}(2)$ group symmetry for $F>2$ achieves exact trajectory-based evolution only when the system Hamiltonian is a linear combination of spin operators \cite{Klimov2002,Klimov2008,Klimov2017}. In contrary, the CMM and wMM methods always satisfy the frozen nuclei limit \cite{He2021a,Liu2021,He2022}. {In this section}, the phase space, phase {trajectory}, and integral {expression} for {TCF}s for time-dependent properties are introduced, {where} the CMM and wMM formulations based on the CPS {are used} as examples.

\section{Phase Space Structure of Finite-State Quantum Systems}\label{sec:phase-space-structure}

The mapping Hamiltonian for the classical mapping model with commutator variables (CMMcv) method {has} been proposed as \cite{He2021}
\begin{equation}
    {H_C(\mathbf{x},\mathbf{p},\mathbf{R},\mathbf{P},\mathbf{\Gamma} ) = \sum_{n,m=1}^{F} \left[\frac{1}{2} \left(x_n+\ii p_n\right)\left(x_m-\ii p_m\right) - \Gamma_{nm} \right] H_{mn}(\mathbf{R},\mathbf{P}),}\label{eq:mapping-Hamiltonian-CMMcv}
\end{equation}
where $\mathbf{\Gamma}$ is a Hermitian matrix whose matrix elements are variables.  {Equation} (\ref{eq:mapping-Hamiltonian-CMMcv}) suggests the following mapping kernel for electronic DOFs,
\begin{equation}
    \hat{K}(\mathbf{x},\mathbf{p},\mathbf{\Gamma}) = \sum_{n,m=1}^{F} \ketbra{n}{m}\left[\frac{1}{2} \left(x_n+\ii p_n\right)\left(x_m-\ii p_m\right) - \Gamma_{nm} \right],\label{eq:mapping-kernel-CMMcv}
\end{equation}
which implies {generalized} phase space formalisms for finite-state quantum systems based on phase space variables $(\mathbf{x},\mathbf{p},\mathbf{\Gamma})$.

{It is heuristic to compare} the {constraint phase space formulation} to the geometric quantum mechanics \cite{STROCCHI1966,Kibble1979,Heslot1985,Brody2001}. Geometric quantum mechanics regards the set of all pure states $\ket{\psi(\mathbf{X})}$ of a quantum system as a classical phase space, while the value of the classical Hamiltonian at {a} phase point $\mathbf{X}$ is defined as $H(\mathbf{X})=\bra{\psi(\mathbf{X})}\hat{H}\ket{\psi(\mathbf{X})}$. The corresponding mapping kernel takes the form 
\begin{equation}
    \hat{K}_{\mathrm{GQM}}(\mathbf{X}) = \ketbra{\psi(\mathbf{X})}{\psi(\mathbf{X})},
\end{equation}
which is the density operator (matrix) of the pure state $\ket{\psi(\mathbf{X})}$  whose matrix rank is $1$. On the other hand, the mapping kernel of the CPS formulation, Eq. (\ref{eq:mapping-kernel-CMM}), can be reorganized as 
\begin{equation}
    \hat{K}(\mathbf{X};\gamma) = \ketbra{\mathbf{X}}{\mathbf{X}} - \gamma \hat{I}, \quad \ket{\mathbf{X}} = \frac{1}{\sqrt{2}}\sum_{n=1}^{F} \left(x_n+\ii p_n\right)\ket{n}. \label{eq:kernel-cmm-gm}
\end{equation}
Since the mapping kernel satisfies $\Tr \hat{K}(\mathbf{X})= 1$, the mapping kernel can be interpreted as a generalized\footnote{The mapping kernel may have negative eigenvalues, while those of a density matrix are always positive semi-definite.} density operator (matrix). Eq. (\ref{eq:kernel-cmm-gm}) is composed of a rank-$1$ part and an identity matrix part. The mapping kernel Eq. (\ref{eq:mapping-kernel-CMMcv}), in contrary, usually has higher rank parts.

We define the phase trajectory in the following way: when the phase point evolves from $\mathbf{X}_0$ to $\mathbf{X}_t$, the mapping kernel {corresponds to the} evolved phase point, $\hat{K}(\mathbf{X}_t)$, is isomorphic to the quantum evolution, i.e.
\begin{equation}
    {\hat{K}(\mathbf{X}_t) = \hat{U}(t) \hat{K}(\mathbf{X}_0) \hat{U}^\dagger(t).} \label{eq:time-covariance}
\end{equation}
All $\hat{U}(t)$ form the unitary group $\mathrm{U}(F)$, which is the group of all $F\times F$ unitary matrices with matrix multiplication as the group operation \cite{Curtis1984}. The condition Eq. (\ref{eq:time-covariance}) is the $\mathrm{U}(F)$ covariance condition for the phase space \cite{Varilly1989a,Varilly1989,Brif1999},
\begin{equation}
    \forall g\in \mathrm{U}(F), \forall \mathbf{X} \in M_{\mathrm{PS}}, \hat{K}(g\cdot \mathbf{X}) = g\hat{K}(\mathbf{X})g^{-1}. \label{eq:covariance-condition}
\end{equation}
In Eq. (\ref{eq:covariance-condition}), two multiplication operations are involved: the multiplication of matrices\footnote{In this section, the matrix form of an operator in the natural basis is considered identical to the operator itself.} and the action of the group $\mathrm{U}(F)$ on the phase space $M_{\mathrm{PS}}$. {The phase space is defined to be ``complete": For any phase point $\mathbf{X}\in M_{\mathrm{PS}}$ and any group element $g\in \mathrm{U}(F)$, the group element $g$ acts on the phase point $\mathbf{X}$ and transforms it to another phase point $g\cdot \mathbf{X}\in M_{\mathrm{PS}}$, i.e. the phase space is the envelope of all trajectories.} While the specific phase space is yet to be determined, the crucial point is the ``completeness'' discussed above, which is the \textit{existence} of a {$\mathrm{U}(F)$} action on the phase space. The constraint phase space satisfies this property, while one cannot self-consistently define such $\mathrm{U}(F)$ action on some other kinds of phase space (e.g., the phase space representation based on SU(2) group symmetry for $F>2$ \cite{Stratonovich1956,Varilly1989,Amiet1991,Amiet2000,Dowling1994,Klimov2002,Klimov2008,Klimov2017}).

\begin{figure}[t]

    {\includegraphics[width=1.0\textwidth]{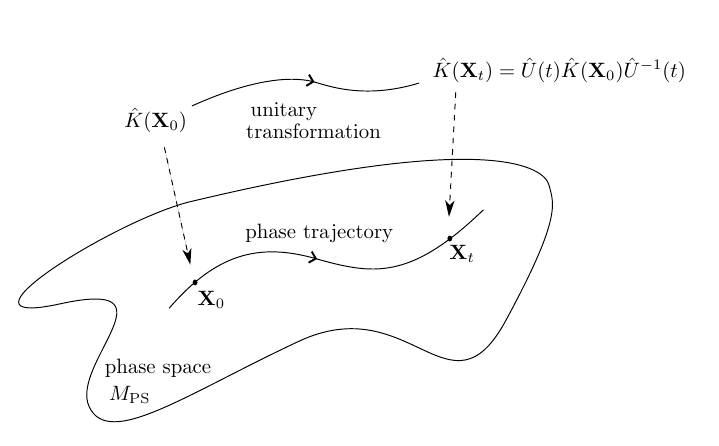}}
    \caption{The phase space structure is determined by the corresponding mapping kernels. Beginning at point $\mathbf{X}_0$ { on} the phase space, the associated mapping kernel $\hat{K}(\mathbf{X}_0)$ {is transformed to} $\hat{K}(\mathbf{X}_t)=\hat{U}(t)\hat{K}(\mathbf{X}_0)\hat{U}^{-1}(t)$ through the unitary time evolution operator $\hat{U}(t)$. The phase trajectory connecting $\mathbf{X}_0$ and $\mathbf{X}_t$ corresponds to the unitary transformation.
}
    \label{fig:phase-space-kernel}
\end{figure}

With the $\mathrm{U}(F)$ covariance condition, the structure of the phase space is entirely determined by the mapping kernel. We first discuss the phase space structure where {we have the} one-to-one correspondence between $\mathbf{X}$ and $\hat{K}(\mathbf{X})$. Each time evolution defines a continuous path {$\{ \hat{U}(\tau) \hat{K}(\mathbf{X}_0) \hat{U}^\dagger(\tau), 0\leq \tau \leq t\}$} of the mapping kernel. With the covariance condition, there is a corresponding trajectory on the phase space $\{ \mathbf{X}_\tau, 0\leq \tau \leq t\}$, {which} describes the dynamics of this time evolution (see Fig. \ref{fig:phase-space-kernel}).

The mapping kernels are Hermitian matrices \cite{He2021}, which can always be diagonalized to be $g_0^{-1}(\mathbf{X})\hat{K}(\mathbf{X}) g_0(\mathbf{X})=\hat{K}_{\mathrm{diag}}$, where $g_0(\mathbf{X})$ {are} unitary {matrices} and $\hat{K}_{\mathrm{diag}}$ {associated} mapping {kernels}. 

Each connected component of the phase space is an orbit under the group action, whose geometric structure is the quotient space $ \mathrm{U}(F)/W$  where $ W$ is the isometry group of any point on the orbit\footnote{The isometry groups of all points on an orbit are isomorphic to {one another.}} \cite{Kirillov1976}. The most straightforward way of obtaining $ W$ is examining all the $\mathrm{U}(F)$ group elements which keep a point on the phase space invariant. It is convenient to use the diagonal mapping kernel $\hat{K}_{\mathrm{diag}}$ as the base point, which must also correspond to a point $\mathbf{X}_{\mathrm{diag}}$ on the phase space. $\hat{K}_{\mathrm{diag}}$ is given by

\begin{equation}
    \begin{split}
        \hat{K}_{\mathrm{diag}} = & \mathrm{diag}\{\lambda^{(1)},\lambda^{(2)},\dots, \lambda^{(F)}\}
        \\= & \mathrm{diag} \{\underbrace{\lambda_1,\dots, \lambda_1}_{d_1}, \underbrace{\lambda_2, \dots, \lambda_2}_{d_2}, \dots, \underbrace{\lambda_l, \dots, \lambda_l}_{d_l}\},
    \end{split}
\end{equation}
where the degeneracy degrees \(d_j\) satisfy \(d_1 \le d_2 \le \dots \le d_l\), \(\sum_{j=1}^l d_j = F\), and \(l \ge 2\). {Under} this circumstance, the isometry group $W$ consists of all the $g_{\mathrm{iso}} \in \mathrm{U}(F)$ so that $g_{\mathrm{iso}}\hat{K}_\mathrm{diag}g_{\mathrm{iso}}^{-1}=\hat{K}_\mathrm{diag}$. Noticing that $\hat{K}_\mathrm{diag}$ is made up of blocks of $\lambda_j \hat{I}_{d_j\times d_j}$, the isometry group $H$ associated with this mapping kernel is isomorphic to \(\mathrm{U}(d_1)\times \mathrm{U}(d_2)\times \dots \times \mathrm{U}(d_l)\), whose elements take the form:
\begin{equation}
    {g_{\mathrm{iso}}} = \left( \begin{array}{cccc}
        (g_1)_{d_1\times d_1} & 0 & \dots & 0 \\
        0 & (g_2)_{d_2\times d_2} & \dots & 0 \\
        \vdots & \vdots & \ddots & \vdots \\
        0 & 0 & 0 & (g_l)_{d_l\times d_l}
    \end{array} \right), \quad g_j\in \mathrm{U}(d_j).
\end{equation}
The structure of such connected component of the phase space is then the quotient space $\mathrm{U}(F) / \prod_{i=1}^{l} \mathrm{U}(d_i)$ \cite{Tilma2004}. As a special case, {when all eigenvalues dergenerate except for one distinct eigenvalue, the phase space of $\mathrm{U}(F)$ coherent states\cite{Tilma2012} is $\mathrm{U}(F) / (\mathrm{U}(1) \times \mathrm{U}(F-1)) \cong \mathrm{SU}(F) / \mathrm{U}(F-1)$,} which is widely explored in geometric quantum mechanics \cite{Heslot1985,Brody2001}. For two-state systems, the quotient space is $\mathrm{U}(2) / (\mathrm{U}(1) \times \mathrm{U}(1)) \cong S^2$, the {two-dimensional} sphere which is the $\mathrm{SU}(2)$ Stratonovich-type phase space \cite{Stratonovich1956,Varilly1989,Amiet1991,Amiet2000,Dowling1994,Klimov2002,Klimov2008,Klimov2017}.

The topology inherent in $\mathrm{U}(F) / \prod_{i=1}^{l} \mathrm{U}(d_i)$ complicates its parameterization. {For example, there are singularities within the complicated EOMs of angles on $\mathrm{SU}(F) / \mathrm{U}(F-1)$, as} discussed in Section 3 of the supplementary material of ref.  \cite{He2022}. It is better to embed $\mathrm{U}(F) / \prod_{i=1}^{l} \mathrm{U}(d_i)$ into $\mathrm{U}(F)/\mathrm{U}(F-r)$, the \textit{complex Stiefel manifolds} \cite{Atiyah1960,Hatcher2002}, which are also denoted as $V_r(\mathbb{C}^F)$ throughout this paper. { Setting $r$ to be any $d_i$ is allowed, and we choose}
\begin{equation}
{ r=F-\max\{d_i\},} \label{eq:r-set}
\end{equation}
{ which is the one requiring least variables.} The parameterization of $\mathrm{U}(F)/\mathrm{U}(F-r)$ directly utilizes the eigenvectors of the mapping kernel, which provides a more practicable way of representing the dynamics on the phase space.

{Note that} $g_0^\dagger(\mathbf{X}) \hat{K}(\mathbf{X}) g_0 (\mathbf{X}) =\hat{K}_{\mathrm{diag}}$ {can be reformulated to} $\hat{K}(\mathbf{X})=g_0(\mathbf{X}) \hat{K}_{\mathrm{diag}}g_0^\dagger(\mathbf{X})$. The $i$-th column of $g_0(\mathbf{X})$ is a normalized eigenvector $\mathbf{v}^{(i)}$ associated with the $i$-th eigenvalue $\lambda^{(i)}$. That is, $g_0(\mathbf{X})=(\mathbf{v}^{(1)},\cdots,\mathbf{v}^{(F)})$, and $\hat{K}(\mathbf{X})=\sum_{i=1}^F\lambda^{(i)}\mathbf{v}^{(i)}{\mathbf{v}^{(i)}}^\dagger$. Denote the eigenvalue of the maximal degeneracy as $-\gamma$. Then, it is convenient to rewrite $\hat{K}(\mathbf{X})$ as
\begin{equation}
   {\hat{K}(\mathbf{X}) = \sum_{i=1}^{r} \left[(\lambda^{(i)}+\gamma) \mathbf{v}^{(i)} {\mathbf{v}^{(i)}}^\dagger\right] - \gamma \hat{I}.}
\end{equation}
Only $r$ eigenvectors are now used for the parameterization. To argue that the $\mathrm{U}(F)/\mathrm{U}(F-r)$ structure exists, consider the isometry subgroup that keeps $(\mathbf{v}^{(1)},\cdots,\mathbf{v}^{(r)})$ invariant, whose group elements are all $\tilde{g}_{\mathrm{iso}}$ such that $\tilde{g}_{\mathrm{iso}}(\mathbf{v}^{(1)},\cdots,\mathbf{v}^{(r)})=(\mathbf{v}^{(1)},\cdots,\mathbf{v}^{(r)})$. We choose a specific point, where
\begin{equation}
    \mathbf{v}^{(i)}_0 = (0, \dots, 0, \underset{i\text{-th}}{1}, 0, \dots, 0)^\mathrm{T}, \quad \text{for}\; i=1,\dots, r,
\end{equation}
{then any $\tilde{g}_{\mathrm{iso}}$ is of the following form,}
\begin{equation}
    {\tilde{g}_{\mathrm{iso}}}=\left(\begin{array}{cc}
        1_{r\times r} & 0 \\
        0 & g_{(F-r)\times (F-r)}
    \end{array}\right), \quad g_{(F-r)\times (F-r)}\in \mathrm{U}(F-r).
\end{equation}
Therefore, the structure of the variable space in this parameterization is the quotient space $\mathrm{U}(F) / \mathrm{U}(F-r)$ \cite{Nakahara2003}, which is the complex Stiefel manifold $V_r(\mathbb{C}^F)$ \cite{Hatcher2002}, the collection of all {$r$-frames $(\mathbf{v}^{(1)},\cdots,\mathbf{v}^{(r)})$} in $\mathbb{C}^F$. The dimension of this manifold is $\dim \mathrm{U}(F) - \dim \mathrm{U}(F-r) = F^2 - (F-r)^2 = 2Fr-r^2$, which is larger than that of $\mathrm{U}(F) / \prod_{i=1}^{l} \mathrm{U}(d_i)$. The difference between the two kinds of manifolds arises from both the global phases of the eigenvectors and the unitary rotations within the subspace composed of eigenvectors with the same eigenvalue.

\begin{figure}[H]
    \centering
    {\includegraphics[width=1.0\textwidth]{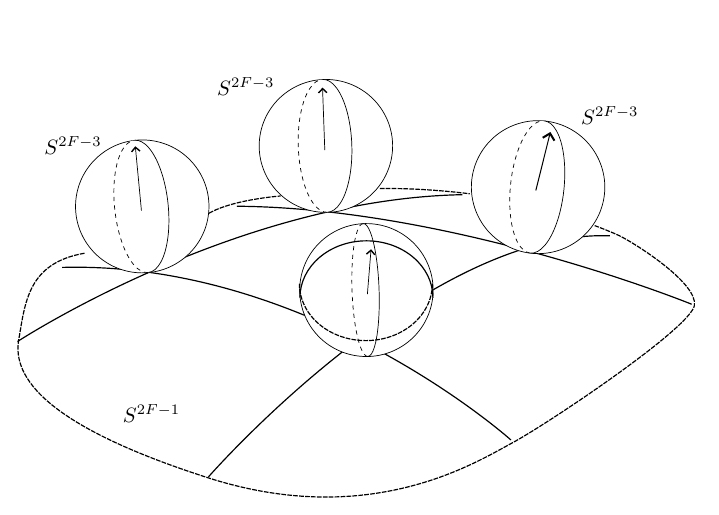}}
    \caption{The complex Stiefel manifold $V_2(\mathbb{C}^F)$ is depicted, where each point on the primary sphere $S^{2F-1}$ is associated with an attached sphere $S^{2F-3}$. These attached spheres are in a non-parallel, twisted configuration, emphasizing the intricate structure of the manifold. {Similar patterns also apply to higher order complex Stiefel manifolds, $V_r(\mathbb{C}^F)$.}}
    \label{fig:stiefel-mfd}
\end{figure}

For the special case of $r=1$, which can be related to the Meyer-Miller Hamiltonian, the complex Stiefel manifold $V_1(\mathbb{C}^F)$ is the sphere $S^{2F-1}$ generated by the eigenvector $\mathbf{v}^{(1)}$. For $r=2$ (see Appendix B for further discussion related to {the first mapping Hamiltonian model of ref. \cite{Liu2016}}), the second eigenvector $\mathbf{v}^{(2)}$ must be orthogonal to the first one. Given a fixed $\mathbf{v}^{(1)}$, all normalized $\mathbf{v}^{(2)}$ orthogonal to $\mathbf{v}^{(1)}$ form the sphere $S^{2F-3}$. The complex Stiefel manifold $V_2(\mathbb{C}^F)$ then can be {interpreted} as attaching a sphere $S^{2F-3}$ to each point of the sphere $S^{2F-1}$, which can be written as $S^{2F-1} \ltimes S^{2F-3}$. The symbol $\ltimes$ is different from the symbol $\times$, the direct product of two spaces. In the latter one, the spheres attached to different points are parallel with each other, while in the former one, there is a ``twist", as shown in Figure \ref{fig:stiefel-mfd}. {Similarly}, a complex Stiefel manifold $V_r(\mathbb{C}^F)$ can be alternatively written as
\begin{equation}
    S^{2F-1} \ltimes S^{2F-3} \ltimes \dots \ltimes S^{2F-2r+1}.
\end{equation}

To conclude this section, we interpret the complex Stiefel manifold as the generalized CPS. For $r=1$, we denote $\mathbf{v}\equiv \mathbf{v}^{(1)}$ throughout this paper. The mapping kernel takes the form:
\begin{equation}
    \hat{K}(\mathbf{X}) = (\lambda^{(1)}+\gamma) \mathbf{v} {\mathbf{v}}^\dagger - \gamma \hat{I}. \label{eq:stiefel-1}
\end{equation}
By assigning $\mathbf{x}+\ii \mathbf{p}=\sqrt{2|\lambda^{(1)}+\gamma|}\mathbf{v}$, Eq. (\ref{eq:stiefel-1}) {becomes}
\begin{equation}
    \hat{K}(\mathbf{X}) = \sum_{n,m=1}^{F}\ketbra{n}{m} \left[\frac{\mathrm{sgn}(\lambda^{(1)}+\gamma)}{2} \left(x_n+\ii p_n\right)\left(x_m-\ii p_m\right) - \gamma \delta_{nm}\right],
\end{equation}
which is reminiscent {of} Eq. (\ref{eq:mapping-kernel-CMM}). The {physical} constraint of the {CPS} variables {is related to} the normalization of the eigenvector $\mathbf{v}$.

For the general {case of} $r>1$,
\begin{equation}
    {\mathbf{x}^{(i)}+\ii \mathbf{p}^{(i)}=\sqrt{2|\lambda^{(i)}+\gamma|}\mathbf{v}^{(i)}}, \quad i=1,\dots, r
\end{equation}
the mapping kernel reads
\begin{equation}
    \hat{K}(\mathbf{X}) = \sum_{n,m=1}^{F}\ketbra{n}{m} \left[\sum_{i=1}^{r}\frac{s^{(i)}}{2} \left(x_n^{(i)}+\ii p_n^{(i)}\right)\left(x_m^{(i)}-\ii p_m^{(i)}\right) - \gamma \delta_{nm}\right],
\end{equation}
where the sign factor of the $i$-th eigenvalue is {\cite{He2021}}
\begin{equation}
    s^{(i)} = \mathrm{sgn}(\lambda^{(i)}+\gamma).\label{eq:gCPS-kernel-sign-factor}
\end{equation}
The relations of the mapping variables come from the normalization and orthogonality of the eigenvectors of the mapping kernel, which read
\begin{equation}
    \begin{split}
        &\sum_{n=1}^{F} \frac{\left(x_n^{(i)}\right)^2+\left(p_n^{(i)}\right)^2}{2}=\left|\lambda^{(i)} + \gamma \right|, \\ &\sum_{n=1}^{F} \left(x_n^{(i)}x_n^{(j)}+p_n^{(i)}p_n^{(j)}\right)=\sum_{n=1}^{F} \left(x_n^{(i)}p_n^{(j)}-p_n^{(i)}x_n^{(j)}\right)=0, \quad i\ne j.
    \end{split}
\end{equation}
These relations are always {satisfied} during the dynamics. For $r>1$, denote $\bar{\mathbf{x}}=\{\mathbf{x}^{(1)},\cdots,\mathbf{x}^{(r)}\}$ and $\bar{\mathbf{p}}=\{\mathbf{p}^{(1)},\cdots,\mathbf{p}^{(r)}\}$. The constraint for a phase space component (diffeomorphic to the complex Stiefel manifold) then reads
\begin{equation}
    \begin{split}
        &{\mathcal{S}_r(\bar{\mathbf{x}},\bar{\mathbf{p}};\boldsymbol{\lambda}) = \prod_{i=1}^{r} \delta\left(\sum_{n=1}^{F} \frac{\left(x_n^{(i)}\right)^2+\left(p_n^{(i)}\right)^2}{2} - \left|\lambda^{(i)} + \gamma \right|\right)} \\ {\times} &{\prod_{i,j=1\atop i<j}^{r}\left[\delta\left(\sum_{n=1}^{F} \left(x_n^{(i)}x_n^{(j)}+p_n^{(i)}p_n^{(j)}\right)\right)\delta\left(\sum_{n=1}^{F} \left(x_n^{(i)}p_n^{(j)}-p_n^{(i)}x_n^{(j)}\right)\right)\right],} \\
    \end{split}\label{eq:constraint-stiefel-mfd}
\end{equation}
and the integral on a component of the generalized CPS is defined as
\begin{equation}
    {\int_{\mathcal{S}_r(\bar{\mathbf{x}},\bar{\mathbf{p}};\boldsymbol{\lambda})}\dd\mu(\bar{\mathbf{x}},\bar{\mathbf{p}})g(\bar{\mathbf{x}},\bar{\mathbf{p}})=\frac{1}{\Omega_r(\boldsymbol{\lambda})}\int F\dd\bar{\mathbf{x}}\dd\bar{\mathbf{p}}\mathcal{S}_r(\bar{\mathbf{x}},\bar{\mathbf{p}};\boldsymbol{\lambda})g(\bar{\mathbf{x}},\bar{\mathbf{p}})}\label{eq:int-generalized-eCMM}
\end{equation}
where
\begin{equation}
    {\Omega_r(\boldsymbol{\lambda})=\int \dd\bar{\mathbf{x}}\dd\bar{\mathbf{p}}\mathcal{S}_r(\mathbf{x},\mathbf{p};\boldsymbol{\lambda}).}
\end{equation}
Similar to the $r=1$ case, integrals on different $\mathcal{S}_r(\bar{\mathbf{x}},\bar{\mathbf{p}};\boldsymbol{\lambda})$ can be joined together by a \textit{quasi}-probability distribution.

Besides, a symplectic structure {intrinsic in} the EOMs of the trajectories on the generalized CPS is discussed in Appendix A. With the parameterization $\{\bar{\mathbf{x}},\bar{\mathbf{p}}\}$, the EOMs of the phase trajectories can be written as
\begin{equation}
    \dot{x}_n^{(k)} = s^{(k)} \frac{\partial H_C}{\partial p_n^{(k)}}, \quad \dot{p}_n^{(k)} = -s^{(k)} \frac{\partial H_C}{\partial x_n^{(k)}}, \quad n=1, \dots, F, \quad k=1,\dots, r, \label{eq:constraint-phase-space-eom}
\end{equation}
which exactly recover the Hamilton's EOMs Eq. (\ref{eq:EOM}) \cite{Meyer1979,Sun1997,Sun1998} for the case \(r=1\) and \(s^{(1)}=1\).

\section{Expressions of Phase Space {Formulations}}\label{sec:correlation-function}

In Section \ref{sec:phase-space-structure}, we {have} discussed the phase space structure and the EOMs of trajectories on the generalized CPS. In this section, we {focus} on the integral expression for {TCF}s. {The general integral expression of TCF is proposed to be \cite{He2024}}
\begin{equation}
    {\Tr\left[\ketbra{n}{m}\hat{U}^\dagger(t) \ketbra{k}{l} \hat{U}(t)\right] \mapsto \frac{1}{\bar{C}_{nm,kl}(t)} \int_{\mathcal{M}_{\mathrm{PS}}} \dd\mu(\mathbf{X}_0) \bar{\mathcal{Q}}_{nm,kl}(\mathbf{X}_0;\mathbf{X}_t),} \label{eq:general-correlation}
\end{equation}
{ where $\bar{C}_{nm,kl}(t)$ can be a time-dependent normalization factor. }

\subsection{Two approaches with covariant kernels naturally leading to exact {dynamics of pure finite-state systems}}

{We first discuss the two classes of phase space mapping formalisms with separable correlation functions that \textit{naturally} lead to the frozen nuclei limit, as pointed out in our previous work \cite{Shang2022thesis} in 2022.} {In these two classes, $\bar{\mathcal{Q}}_{nm,kl}(\mathbf{X}_0;\mathbf{X}_t)$ can always be separated as}
\begin{equation}
    \bar{\mathcal{Q}}_{nm,kl}(\mathbf{X}_0; \mathbf{X}_t) = \left[\hat{K}_\rho(\mathbf{X}_0)\right]_{mn} \left[\hat{K}_A(\mathbf{X}_t)\right]_{lk},
\end{equation}
{and $\bar{C}_{nm,kl}(t)$ is a time-independent constant which is set to be 1 in this subsection.} The first class utilizes a covariant observable kernel $\hat{K}_A(\mathbf{X})$ while the second class features a covariant density kernel $\hat{K}_\rho(\mathbf{X})$. In this subsection, all {TCF}s are expressed by Eq. (\ref{eq:TD-expectation-value}).

In the first class, the observable kernel $\hat{K}_A(\mathbf{X})$ is set to be covariant, the physical quantity estimator $\tilde{A}_C(\mathbf{X})$ satisfies the following property:
\begin{equation}
    \begin{split}
    {\tilde{A}_C(\mathbf{X}_t) }&{= \mathrm{Tr}\left[\hat{A} \hat{U} (t)\hat{K}_A(\mathbf{X}_0) \hat{U}^\dagger(t)\right] }\\
    &{= \Tr\left[\hat{U}^\dagger (t)\hat{A} \hat{U}(t) \hat{K}_A(\mathbf{X}_0)\right]} \\ 
    &{= \Tr\left[\hat{A}(t)\hat{K}_A(\mathbf{X}_0)\right] =:\tilde{A}_C(\mathbf{X}_0;t),} \\ 
    \end{split}
\end{equation}
Therefore, utilizing Eq. (\ref{eq:TI-expectation-value}), the integral expression {of} Eq. (\ref{eq:TD-expectation-value}) is exact in the frozen nuclei limit,
\begin{equation}
    \begin{split}
    & {\int_{\mathcal{M}_{\mathrm{PS}}} \dd\mu(\mathbf{X}_0) \rho_C(\mathbf{X}_0) \tilde{A}_C(\mathbf{X}_t)} \\
    = & {\int_{\mathcal{M}_{\mathrm{PS}}} \dd\mu(\mathbf{X}_0) \rho_C(\mathbf{X}_0) \tilde{A}_C(\mathbf{X}_0;t)=\mathrm{Tr}\left[\hat{\rho}\hat{A}(t)\right],} \\
    \end{split}
\end{equation}
as long as the exact mapping condition Eq. (\ref{eq:TI-expectation-value}) holds, regardless of the form of the density kernel $\hat{K}_{\rho}(\mathbf{X})$.

In the second class, where the density kernel $\hat{K}_\rho(\mathbf{X})$ is set to be covariant. Let $\mathbf{X}_t = \mathbf{Y}_{0}$, then
\begin{equation}
    \begin{split}
    &{\int_{\mathcal{M}_{\mathrm{PS}}} \dd\mu(\mathbf{X}_0) \rho_C(\mathbf{X}_0) \tilde{A}_C(\mathbf{X}_t)} \\
    =&{\int_{\mathcal{M}_{\mathrm{PS}}} \dd\mu(\mathbf{Y}_{-t}) \rho_C(\mathbf{Y}_{-t}) \tilde{A}_C(\mathbf{Y}_0).} \\ 
    \end{split} \label{eq:time-translation}
\end{equation}
Since $\dd \mu$ is the \textit{invariant measure},  $\dd\mu(\mathbf{X}_0) = \dd \mu(\mathbf{X}_t)$ so that $\dd\mu(\mathbf{Y}_{-t}) = \dd \mu(\mathbf{Y}_0)$. Also, as $\hat{K}_\rho(\mathbf{X})$ is covariant,
\begin{equation}
    \begin{split}
    {\rho_C(\mathbf{Y}_{-t})}&{=\Tr\left[\hat{\rho}\hat{U}^\dagger (t)\hat{K}_\rho(\mathbf{Y}_0) \hat{U}(t)\right]}\\
    &{=\Tr\left[\hat{U}(t)\hat{\rho}\hat{U}^\dagger (t)\hat{K}_\rho(\mathbf{Y}_0) \right]}\\
    &{=:\Tr\left[\hat{\rho}(t)\hat{K}_\rho(\mathbf{Y}_0) \right]=\rho_C(\mathbf{Y}_0;t).}\\
    \end{split}
\end{equation}
Note that the definition $\hat{\rho}(t)=\hat{U}(t)\hat{\rho}\hat{U}^\dagger(t)$ is different from $\hat{A}(t)=\hat{U}^\dagger (t)\hat{A} \hat{U}(t)$, so that $\Tr\left[\hat{\rho}(t)\hat{A}\right]=\Tr\left[\hat{\rho}\hat{A}(t)\right]$. { By} substituting the arguments above into Eq. (\ref{eq:time-translation}), we obtain
\begin{equation}
    \begin{split}
    & {\int_{\mathcal{M}_{\mathrm{PS}}} \dd\mu(\mathbf{X}_0) \rho_C(\mathbf{X}_0) \tilde{A}_C(\mathbf{X}_t)} \\
    = & {\int_{\mathcal{M}_{\mathrm{PS}}} \dd\mu(\mathbf{Y}_0) \rho_C(\mathbf{Y}_0;t) \tilde{A}_C(\mathbf{Y}_0)} \\
    = & {\Tr\left[\hat{\rho}(t)\hat{A}\right]=\Tr\left[\hat{\rho}\hat{A}(t)\right],} \\
    \end{split}
\end{equation}
which also leads to the exact frozen nuclei limit when the exact mapping condition holds, { irrespective of the form of the observable kernel $\hat{K}_A(\mathbf{X})$.}

Thus, in the two classes of mapping formalisms, we have great flexibility in choosing the function type of the kernels. In this paper, we label the mapping formalisms with the charactistics of the density kernel and the observable kernel: e.g. when both the density kernel and the observable kernel are covariant, we label the mapping formalism as the covariant-covariant (cc) type. Here, we consider the covariant-covariant (cc) type, {covariant-noncovariant (cx) type, and noncovariant-covariant (xc) type} {{TCF}s}. The examples discussed in this subsection are briefly summarized in Table \ref{tab:correlation-function}. The derivation details of the exact mapping condition leading to the frozen nuclei limit of these examples are demonstrated in Appendix C.

\begin{sidewaystable}[thp]
    \centering
    \begin{threeparttable}
    \resizebox{\linewidth}{!}{
        \begin{tabular}{ccccccc}
            \toprule
            type & method & phase space & $\hat{K}_\rho$& characteristic & $\hat{K}_A$&characteristic \\
            \midrule
            cc & CMM \cite{He2019,He2021a}\tnote{1} & $V_1(\mathbb{C}^F)$ & Eq.(\ref{eq:mapping-kernel-CMM})  & $\gamma\in(-1/F,    +\infty)$   & Eq.(\ref{eq:inverse-mapping-kernel-CMM})  &   \\

            cc & wMM \cite{He2022} & multiple $V_1(\mathbb{C}^F)$ & Eq.(\ref{eq:mapping-kernel-CMM}) &  $\gamma\in(-1/F,    +\infty)$  & Eq.(\ref{eq:mapping-kernel-CMM}) &  \\

            cx & cornered simplex & $V_1(\mathbb{C}^F)$ & Eq.(\ref{eq:mapping-kernel-CMM}) & $\gamma\in(0,+\infty)$  &  Eq.(\ref{eq:window-function-cw2}) (diag) & $\mathrm{w}_n=h(e_n-1)$ \\

            xc & triangle \cite{Cotton2016,Cotton2019,He2024}  & multiple $V_1(\mathbb{C}^F)$ & Eq.(\ref{eq:triangle-window-Krho}) &  triangle region  & Eq.(\ref{eq:mapping-kernel-CMM}) &  \\

            xc & Ehrenfest \cite{Ehrenfest1927,Delos1972,Billing1975,Micha1983} & $V_1(\mathbb{C}^F)$ & $\begin{cases}
            \text{Eq.(\ref{eq:Ehr-Krho-diag}) (diag)} \\ \text{Eq.(\ref{eq:Ehrenfest-Krho-offdiag}) (off-diag)\tnote{2}}
            \end{cases}$ & & Eq.(\ref{eq:mapping-kernel-CMM})  & $\gamma=0$ \\

            xc & $\Lambda$-point\tnote{2} & $V_1(\mathbb{C}^F)$ & $\begin{cases}
            \text{Eq.(\ref{eq:focused-Krho-diag}) (diag)} \\ \text{Eq.(\ref{eq:focused-Krho-offdiag}) (off-diag)\tnote{2}}
            \end{cases}$ &  &  Eq.(\ref{eq:mapping-kernel-CMM}) & $\gamma\in(0,+\infty)$ \\

            xc & DTWA \cite{Blakie2008,Polkovnikov2010}($F=2$) & $V_1(\mathbb{C}^F)$ & Eq.(\ref{eq:Krho-GDTWA})\tnote{2}  & & Eqs.(\ref{eq:KA-GDTWA},\ref{eq:Krho-GDTWA}) &  $\gamma=\gamma_{\mathrm{w}}$ \\

            xc & GDTWA \cite{Zhu2019,Lang2021}($F\ge 3)$  & $V_2(\mathbb{C}^F)$ & Eq.(\ref{eq:Krho-GDTWA})\tnote{2}  &  &  Eqs.(\ref{eq:KA-GDTWA},\ref{eq:Krho-GDTWA}) & \begin{tabular}{c}$\{(1+\sqrt{2F-1})/2$,\\ $(1-\sqrt{2F-1})/2$,\\ $0, \dots, 0\}$\tnote{3} \cite{Lang2021}\end{tabular}  \\
            \bottomrule
        \end{tabular}
    }
    \scriptsize{
    \begin{tablenotes}
        \item[1] {The P, Q, W versions of Weyl-Stratonovich phase space \cite{Stratonovich1956,Tilma2012} correspond to special cases of CPS formulation in the CMM method.}

        \item[2] We propose new expressions for off-diagonal elements of the density kernel of the Ehrenfest method, the $\Lambda$-point method and the (G)DTWA method. 
        
        Please refer to \textit{The restricted-covariant (rc) type correlation function} part. { The $\Lambda$-point method includes the focused method of ref. \cite{Mannouch2020a}.}
        
        \item[3] The full eigenvalue set is presented here to characterize the covariant mapping kernel on $V_2(\mathbb{C}^F)$.
    \end{tablenotes}
    }
    \end{threeparttable}
    \caption{The classification of covariant correlation functions on the generalized constraint phase space.}
    \label{tab:correlation-function}
\end{sidewaystable}

\subsection*{The covariant-covariant (cc) type correlation function}

The CPS formulation in \textbf{CMM method} \cite{He2019,He2021a} serves as a fundamental example within the generalized CPS framework. This method employs a single complex Stiefel manifold, $V_1(\mathbb{C}^F)$, as the {phase space for electronic DOFs. For the {TCF}, both $\hat{K}_\rho=\hat{K}$ (given by Eq. (\ref{eq:mapping-kernel-CMM})) and $\hat{K}_A$ (given by Eq. (\ref{eq:inverse-mapping-kernel-CMM})) are covariant.} The eigenvalues of the phase space kernel $\hat{K}$ are $\{1+(F-1)\gamma, -\gamma, \dots, -\gamma \}$, determined by the {phase space parameter} $\gamma\in(-1/F, +\infty)$. {In spirit of the unified framework of mapping models with coordinate-momentum variables of ref. \cite{Liu2016}, the CMM methods were firstly proposed and derived in ref. \cite{He2019} in 2019 for composite systems with general $F$-state systems ($F\geq 2$).  The CPS with coordinate-momentum variables with the general value of $\gamma$ was first indicated by Eqs. (1), (5), (7), (19), and (28) of ref. \cite{He2019}, and the CPS with action-angle variables for the $\gamma=0$ case was first presented by Eqs. (A4)-(A5) of Appendix A of ref. \cite{He2019}.  Following refs. \cite{Liu2016, He2019, He2019thesis}, refs. \cite{He2021a, He2021} explicitly showed that the phase space parameter $\gamma$ should lie in a \textit{continuous} range $(-1/F, +\infty)$, and pointed out that the P, W, and Q versions of Stratonovich phase space\cite{Stratonovich1956, Tilma2012, Garcia2019,Berezin1975} are three special cases of constraint coordinate-momentum phase space with $\gamma=1, (\sqrt{1+F}-1)/F$ and $0$ used in CMM first presented in refs. \cite{Liu2016, He2019} for general $F$-state systems (especially for $F\geq 3$).}  In addition to ref. \cite{He2021}, the relation has also been presented in ref. \cite{Liu2021}, Appendix 3 of ref. \cite{He2022}, and ref. \cite{Lang2024}.

The \textbf{classical mapping model with commutator variables (CMMcv)  method} \cite{He2021} employs the CMM {TCF}s, while the phase space is generated by the mapping kernel Eq. (\ref{eq:mapping-kernel-CMMcv}), which is a combination of $V_r(\mathbb{C}^F)$ depending on the initial sampling. Each $r$ corresponding to one set of initial sampled variables can be straightforwardly calculated by Eq. (\ref{eq:r-set}).

The \textbf{wMM method} \cite{He2022} introduces a weighted average over multiple complex Stiefel manifolds $V_1(\mathbb{C}^F)$ characterized by different $\gamma$. Although the exact mapping condition may not be  satisfied on each sphere $V_1(\mathbb{C}^F)$ separately, it holds after applying the weighted average. Various choices for the weight function $w(\gamma)$ are permissible, including a finite sum of Dirac delta functions \cite{He2022}. 

Recently, refs. \cite{Lang2022, Lang2024} {proposed} a criterion for the intra-electron correlation derived from short-time analysis. { It is not difficult to show that {when the observable kernel is covariant}, the intra-electron correlation is intrinsically equivalent to}
\begin{equation}
    \frac{1}{2}\Tr\left[\hat{\rho}\{\hat{A},\hat{H}\}\right] = \int_{\mathcal{M}_{\mathrm{PS}}} \dd\mu(\mathbf{X}) \Tr\left[\hat{\rho}\hat{K}_\rho(\mathbf{X})\right]\Tr\left[\hat{A}\hat{K}_A(\mathbf{X})\right]\Tr\left[\hat{H}\hat{K}(\mathbf{X})\right].\label{eq:intra-electron-correlation}
\end{equation}
{When both $\hat{K}_\rho$ and $\hat{K}_A$ are set to be the covariant kernel $\hat{K}$ as Eq. (\ref{eq:mapping-kernel-CMM}),} equation (\ref{eq:intra-electron-correlation}) holds for wMM if the weight function $w(\gamma)$ satisfies
\begin{equation}
    \int_{-1/F}^{+\infty} w(\gamma)(1+F\gamma)^3 \dd\gamma = \frac{1}{2}(1+F)(2+F).
\end{equation}
This condition improves the short-time behavior of the wMM method \cite{He2022} for the FMO monomer model\cite{Ishizaki2009, He2021, Liu2021, He2024}.

\subsection*{The covariant-noncovariant (cx) type correlation function}

When the density kernel for initial sampling is covariant, it is possible to choose {any function for a matrix element of the observable kernel other than }the covariant case. One particular choice is to use window functions {in} the matrix elements of $\hat{K}_A$, which can be defined on one $V_1(\mathbb{C}^F)$ sphere or multiple $V_1(\mathbb{C}^F)$ spheres. The positive semi-definite window function $\mathrm{w}_n(\mathbf{X})$ for the $n$-th state is typically non-zero in a region around the pole on the phase space corresponding to the $n$-th state, and is zero elsewhere. In the discussions from here to the end of the main text, we do not demand the density kernel and the observable kernel to satisfy $\mathrm{Tr}\left[\hat{K}_\rho(\mathbf{X})\right]=1$ and $\mathrm{Tr}\left[\hat{K}_A(\mathbf{X})\right]=1$.

One example denoted as \textbf{cornered-simplex} can be constructed by considering the window function
\begin{equation}
    {\left[\hat{K}_\rho(\mathbf{x},\mathbf{p})\right]_{nn}}=\mathrm{w}_n(\mathbf{x},\mathbf{p}) = \mathcal{N}^{-1} h\left(\frac{x_n^2+p_n^2}{2}-1\right),\label{eq:window-function-cw2}
\end{equation}where the Heaviside step function $h(z)$ is defined as
\begin{equation}
    h(z) = \begin{cases}
        1, & z\ge 0, \\
        0, & z< 0.
    \end{cases}
\end{equation}
The normalization factor $\mathcal{N}=F\left(\frac{F\gamma}{1+F\gamma}\right)^{F-1}$ is chosen such that
\begin{equation}
    \int_{\mathcal{S}(\mathbf{x},\mathbf{p};\gamma)} \dd\mu(\mathbf{x},\mathbf{p}) \mathrm{w}_n(\mathbf{x},\mathbf{p}) = 1.
\end{equation}
Both the covariant density kernel $\hat{K}_A(\mathbf{x},\mathbf{p})$ and the phase space kernel take the form of Eq. (\ref{eq:mapping-kernel-CMM}), where the factor $\gamma$ can be chosen as any positive real number. It should be noticed that though for this method the normalization factor $\bar{C}_{nm,kl}(t)$ is a constant for pure electronic systems, it would be time-dependent when nuclear DOFs are also involved {in the trajectory-based dynamics}. As $\gamma$ approaches zero, the window function becomes more and more concentrated around the pole corresponding to the $n$-th state, whose limit would be the Dirac delta function corresponding to the Ehrenfest dynamics (see below).

\subsection*{The noncovariant-covariant (xc) type correlation function}

For this kind of correlation function, the observable kernel $\hat{K}_A(\mathbf{X})$ is covariant, while {any function} could be utilized in the density kernel $\hat{K}_\rho(\mathbf{X})$. One example is {by utilizing} the \textbf{triangle window function} proposed by Cotton and Miller, \cite{Cotton2016,Cotton2019}
\begin{equation}
    \mathrm{w}_{n}^{\mathrm{TW}}(\mathbf{x},\mathbf{p}) = h\left(\frac{x_n^2+p_n^2}{2}-1\right) \prod_{m\neq n} h \left(2-\frac{x_n^2+p_n^2}{2}-\frac{x_m^2+p_m^2}{2}\right).
    \label{eq:triangle-window-function}
\end{equation}
The weight function over the multiple spheres, $w_{\mathrm{TW}}(\gamma)$, reads \cite{He2024}
\begin{equation}
    {w_{\mathrm{TW}}(\gamma) = \mathcal{N}_{\mathrm{TW}}\frac{(1+F\gamma)^{F-1}}{(F-1)!} h(\gamma) h(1-F^{-1}-\gamma),} \label{eq:SQC-weight}
\end{equation}
where $\mathcal{N}_{\mathrm{TW}} = (F\cdot F!)/\left(F^F - 1\right)$ so that the integral of the weight function over the $\gamma$ axis is $1$. {The elements of the density kernel can be constructed with triangle window functions as \cite{He2024}} 
\begin{equation}
\begin{split}
{\left[\hat{K}_\rho(\mathbf{x},\mathbf{p})\right]_{nn} }&{= 2\mathcal{N}_{\mathrm{TW}}^{-1}\mathrm{w}_{n}^{\mathrm{TW}}(\mathbf{X})\left(2-\frac{x_n^2+p_n^2}{2}\right)^{2-F},} \\
\left[\hat{K}_\rho(\mathbf{x},\mathbf{p})\right]_{nm} &{=\frac{12}{5}\mathcal{N}_{\mathrm{TW}}^{-1}\sum_{k=n,m}\mathrm{w}_{k}^{\mathrm{TW}}(\mathbf{x},\mathbf{p})\left(2-\frac{x_k^2+p_k^2}{2}\right)^{2-F} }\\ &{\times \frac{1}{2}\left(x_n+\ii p_n\right)\left(x_m-\ii p_m\right),} \\
\end{split}
\label{eq:triangle-window-Krho}
\end{equation}
and the phase space kernel $\hat{K}(\mathbf{x},\mathbf{p})$ is Eq. (\ref{eq:mapping-kernel-CMM}) where the phase space parameter $\gamma$ is $[\sum_{n=1}^F(x_n^2+p_n^2)/2-1]/F$. The observable kernel $\hat{K}_A(\mathbf{x},\mathbf{p})$ can either be set to be the same as the phase space kernel, or, as pointed out by Cotton and Miller \cite{Cotton2019}, to be Eq. (\ref{eq:mapping-kernel-CMM}) with $\gamma\equiv 1/3$ which also satisfies the exact mapping condition. We note that there are alternative approaches for the population-population correlation function utilizing triangle window functions, as will be elaborated in Section \ref{sec:window-window-type-TCF}.

When introducing generalized functions (e.g. Dirac delta function) into $\hat{K}_\rho(\mathbf{X})$, the initial sampling is constrained to lower-dimensional subsets of the (generalized) constraint phase space (sCPS). It should be emphasized that the phase space trajectories starting from sCPS in most time can cover the whole CPS. The first example is \textbf{Ehrenfest dynamics} \cite{Delos1972,Billing1975,Micha1983} in spirit of the renowned Ehrenfest theorem \cite{Ehrenfest1927}, where the phase space sampling on the phase space corresponding to a pure $n$-th electronic state is a single phase point. The diagonal elements of the density kernel corresponding to Ehrenfest dynamics is
\begin{equation}
    {\left[\hat{K}^{\mathrm{Ehr}}_{\rho}(\mathbf{x},\mathbf{p})\right]_{nn} = \mathcal{N}^{-1}_{\mathrm{Ehr}} \prod_{m\neq n} \delta\left(x_m\right)\delta\left(p_m\right),}\label{eq:Ehr-Krho-diag}
\end{equation}
where the normalization factor is $\mathcal{N}_{\mathrm{Ehr}}=2\pi F/\Omega(\gamma=0)$. The observable kernel $\hat{K}_A(\mathbf{x},\mathbf{p})$ is the same as the phase space kernel Eq. (\ref{eq:mapping-kernel-CMM}) with $\gamma=0$. For the population-population and population-coherence correlation functions, Ehrenfest dynamics leads to exact pure electronic dynamics. In this paper, we propose a form of the off-diagonal elements of $\hat{K}^{\mathrm{Ehr}}_\rho(\mathbf{x},\mathbf{p})$:
\begin{equation}
    \begin{split}
        & \left[\hat{K}_\rho^{\mathrm{Ehr}}(\mathbf{x},\mathbf{p})\right]_{nm} \\ =\; & \frac{2\mathcal{N}^{-1}_{\mathrm{Ehr}}}{\pi} \frac{(x_n+\ii p_n)(x_m - \ii p_m)}{2} \times \delta\left(\frac{x_m^2+p_m^2}{2}-\frac{1}{2}\right) \prod_{l\neq \{m,n\}}\delta(x_l)\delta(p_l).
    \end{split}\label{eq:Ehrenfest-Krho-offdiag}
\end{equation}

Another method, usually called the focused method \cite{Mannouch2020a, Bonella2003,Bonella2005}, also employs sCPS for its initial condition. For simplicity, we only discuss the {method} on a single $V_1(\mathbb{C}^F)$ phase space. When projecting CPS to the simplex of action variables, the sCPS used in this method for the population-population correlation functions shrinks to a point on the straight line between the center of simplex and one vertex. {This simplex is a tetrahedron when $F=4$, which can be embedded into a cube. On this embedding, the sampling point lies on the cube diagonal}, which is usually denoted as a $\Lambda$ point in solid state physics \cite{Bouckaert1936} {if one regard this cube as the first Brillouin zone of a cubic crystal}. For this reason, we denote this sampling method as \textbf{$\Lambda$-point}. Using the observable kernel $\hat{K}_A(\mathbf{x},\mathbf{p})$ same as { the phase space kernel} Eq. (\ref{eq:mapping-kernel-CMM}) {on $V_1(\mathbb{C}^F)$ with a positive phase space} parameter $\gamma$, the initial sampling of {the $\Lambda$-point} method for the $n$-th electronic state \cite{Mannouch2020a} is generated by the {density} kernel
\begin{equation}
    \left[\hat{K}^{\mathrm{\Lambda}}_\rho(\mathbf{x},\mathbf{p})\right]_{nn} = {\mathcal{N}^{-1}_{\mathrm{\Lambda}}}\prod_{m\neq n}\delta\left(\frac{x_m^2+p_m^2}{2}-\gamma\right),\label{eq:focused-Krho-diag}
\end{equation}
where the normalization factor ${\mathcal{N}_{\mathrm{\Lambda}}}=F(2\pi)^F/\Omega(\gamma)$  such that 
\begin{equation}
    \int_{\mathcal{S}(\mathbf{x},\mathbf{p};\gamma)} \dd\mu(\mathbf{x},\mathbf{p}) \left[\hat{K}_\rho^{\mathrm{\Lambda}}(\mathbf{x},\mathbf{p})\right]_{nn} = 1.
\end{equation}
The initial condition of off-diagonal elements of the $\Lambda$-point method have been defined in a fashion of traversing the sampling corresponding to all the $F$ electronic states \cite{Mannouch2020a}. In this paper, we propose an alternative method that $\left[\hat{K}_\rho^\Lambda(\mathbf{x},\mathbf{p})\right]_{nm}$ should only involve sampling on the $n$-th state and $m$-th state,
\begin{equation}
    \begin{split}
        \left[\hat{K}_\rho^\Lambda(\mathbf{x},\mathbf{p})\right]_{nm} &= \frac{4{\mathcal{N}^{-1}_{\mathrm{\Lambda}}}}{(1+2\gamma)^2} \frac{(x_n+\ii p_n)(x_m - \ii p_m)}{2}\\ &\times \delta\left(\frac{x_m^2+p_m^2}{2}-\frac{1+2\gamma}{2}\right) \prod_{l\neq \{m,n\}}\delta\left(\frac{x_l^2+p_l^2}{2}-\gamma\right).
    \end{split}\label{eq:focused-Krho-offdiag}
\end{equation}
When $\gamma\rightarrow 0$, the $\Lambda$-point method approaches the Ehrenfest dynamics.

The \textbf{discrete truncated Wigner approximation} (\textbf{DTWA}) \cite{Schachenmayer2015} for a two-state system and its multi-state version \textbf{generalized discrete truncated Wigner approximation} (\textbf{GDTWA}) \cite{Zhu2019,Lang2021} feature sampling on discrete points. Particularly, the DTWA method is related to the discrete Wootters phase space. The (G)DTWA methods lead to rc-type correlation functions on a single complex Stiefel manifold. The phase space corresponding to the DTWA method is $V_1(\mathbb{C}^F)$ with $\gamma=\gamma_\mathrm{w}$, while that of the GDTWA method is $V_2(\mathbb{C}^F)$. For DTWA, this embedding of discrete phase space into continuous phase space is discussed in ref. \cite{Zunkovic2015}. The observable kernel of DTWA is the phase space kernel Eq. (\ref{eq:mapping-kernel-CMM}) with $\gamma=\gamma_\mathrm{w}$, and that of GDTWA reads
\begin{equation}
    \begin{split}
    &{\hat{K}_A^{\mathrm{GDTWA}}(\bar{\mathbf{x}},\bar{\mathbf{p}})}\\
    =&{\sum_{m,n=1}^{F} \left[\frac{1}{2} \left(x_n^{(1)}+\ii p_n^{(1)}\right)\left(x_m^{(1)}-\ii p_m^{(1)}\right) - \frac{1}{2} \left(x_n^{(2)}+\ii p_n^{(2)}\right)\left(x_m^{(2)}-\ii p_m^{(2)}\right) \right]}\\
    &\quad\quad\ketbra{n}{m}.\\
    \end{split}
\end{equation}
When the initial state is the pure $n$-th electronic state, the corresponding diagonal term of the density kernel of DTWA/GDTWA is the summation of delta functions,
\begin{equation}
    {\left[\hat{K}_{\rho}^{\mathrm{(G)DTWA}}(\mathbf{X})\right]_{nn} = \mathcal{N}_{\mathrm{(G)DTWA}}^{-1}\sum_{\alpha_n=1}^{2^{2(F-1)}} \delta(\mathbf{X}-\mathbf{X}_{\alpha_n}),}
\end{equation}
so that
\begin{equation}
    {\left[\hat{K}_{A}^{\mathrm{(G)DTWA}}(\mathbf{X}_{\alpha_n})\right]_{ij}} = 
    \begin{cases}
    &{1,\quad \mathrm{if}\quad i=j=n,}\\
    &{(\delta_i + \ii \sigma_i)/2,\quad \mathrm{if}\quad i\neq n ~ \mathrm{and}~ j = n,}\\
    &{(\delta_j - \ii \sigma_j)/2,\quad \mathrm{if}\quad i = n ~ \mathrm{and}~ j \neq n,}\\
    &{0,\quad\mathrm{otherwise}}\\
    \end{cases} \label{eq:KA-GDTWA}
\end{equation}
for DTWA and GDTWA. Here, the collective index $\alpha_n$ is defined as
\begin{equation}
{\alpha_n=(\delta_1,\cdots,\delta_{n-1},\delta_{n+1},\cdots,\delta_F;\sigma_1,\cdots,\sigma_{n-1},\sigma_{n+1},\cdots,\sigma_F).}
\end{equation}
Each {of} $\delta_i$ and $\sigma_i$ is chosen from $\{-1, 1\}$, thus the total number of $\alpha_n$ is $2^{2(F-1)}$. The {phase space} kernel $\hat{K}^{\mathrm{(G)DTWA}}(\mathbf{X})$ is the same as $\hat{K}^{\mathrm{(G)DTWA}}_A(\mathbf{X})$. The set of eigenvalues of the phase space kernels is \cite{Lang2021}
\begin{equation}
    \{\boldsymbol{\lambda}\}=\left\{\frac{1+\sqrt{2F-1}}{2}, \frac{1-\sqrt{2F-1}}{2}, 0, \dots, 0\right\},\label{eq:eig-GDTWA}
\end{equation}
so that the mathematical structure of phase space is the single complex Stiefel manifold $V_1(\mathbb{C}^F)$ when $F=2$ and $V_2(\mathbb{C}^F)$ when $F>2$, and $\mathcal{N}_{\mathrm{DTWA}}=2^{2(F-1)}F/\Omega(\gamma_\mathrm{w})$ and $\mathcal{N}_{\mathrm{GDTWA}}=2^{2(F-1)}F/\Omega_2(\boldsymbol{\lambda})$. The time evolution of the phase space variables are { governed by} Eq. (\ref{eq:constraint-phase-space-eom}). It is straightforward to relate the GDTWA method to the phase space structure of Model 1 of refs. \cite{Liu2016,He2019} determined by the Hamilton's EOMs governed by the Li-Miller Hamiltonian \cite{Li2012}, as elaborated in Appendix B.

In addition, we can construct the off-diagonal terms $\left[\hat{K}_{\rho}^{\mathrm{(G)DTWA}}(\mathbf{X})\right]_{nm}$ of the density kernel for DTWA and GDTWA, leading to the expression for the total density kernel,
\begin{equation}
    {\hat{K}_\rho^{\mathrm{(G)DTWA}} (\mathbf{X}) = \mathcal{N}_{\mathrm{(G)DTWA}}^{-1} \sum_{\nu=1}^F\sum_{\alpha_\nu=1}^{2^{2(F-1)}} \left[\hat{K}(\mathbf{X}_{\alpha_\nu})\right] \delta(\mathbf{X}-\mathbf{X}_{\alpha_\nu}).}\label{eq:Krho-GDTWA}
\end{equation}
It is easy to observe from Eq. (\ref{eq:Krho-GDTWA}) and Eq. (\ref{eq:KA-GDTWA}) that when $\hat{\rho}=\ketbra{n}{n}$, only the phase points $\mathbf{X}_{\alpha_n}$ corresponding to $n$-th state are involved; and when $\hat{\rho}=\ketbra{n}{m}$, only the phase points $\mathbf{X}_{\alpha_n}$ and $\mathbf{X}_{\alpha_m}$ are involved.

\subsection{{Positive semi-definite population-population correlation functions}}\label{sec:window-window-type-TCF}

When considering the population-population correlation functions, the left hand side of Eq. \eqref{eq:general-correlation}, which is the fully quantum mechanical result, is positive semi-definite. However, the negative values of the covariant kernel often lead to the negative population problem for trajectory-based approximate methods. To avoid this non-physical problem, we introduce the window-window (ww) type correlation functions that satisfy the frozen nuclei limit for the $F=2$ case. Within this regime, the contribution of each trajectory to the integral of the population-population correlation function is positive semi-definite, which excludes the non-physical negative population results in trajectory-based methods for composite systems. 

The population-population correlation function with \textbf{triangle window functions} used for the symmetric quasiclassical (SQC) method \cite{Cotton2016,Cotton2019} is constructed on multiple $V_1(\mathbb{C}^F)$. The weight function $w_{\mathrm{TW}}(\gamma)$ is the same as Eq. (\ref{eq:SQC-weight}), and the diagonal elements of the density kernel are the same as those defined in Eq. (\ref{eq:triangle-window-Krho}). The diagonal elements of the observable kernel are
\begin{equation}
    {\left[\hat{K}_A^{\mathrm{TW}}(\mathbf{x},\mathbf{p})\right]_{nn} = h\left(\frac{x_n^2+p_n^2}{2}-1\right) \prod_{m\neq n} h \left(1-\frac{x_m^2+p_m^2}{2}\right).}
\end{equation}
In the expression { of} Eq. (\ref{eq:general-correlation}),
\begin{equation}
{\bar{\mathcal{Q}}_{nn,mm}^{\mathrm{TW}}(\mathbf{x}_0,\mathbf{p}_0;\mathbf{x}_t,\mathbf{p}_t)=\left[\hat{K}_\rho^{\mathrm{TW}}(\mathbf{x}_0,\mathbf{p}_0)\right]_{nn}\left[\hat{K}_A^{\mathrm{TW}}(\mathbf{x}_t,\mathbf{p}_t)\right]_{mm}},
\end{equation}
{ In addition,} the time-dependent normalization factor reads
\begin{equation}
    {\bar{C}_{nn,mm}^{\mathrm{TW}}(t)=\sum_{k=1}^F \int_{0}^{1-1/F}w(\gamma)\int_{\mathcal{S}(\mathbf{x}_0,\mathbf{p}_0;\gamma)}\dd\mu(\mathbf{x}_0,\mathbf{p}_0) \bar{\mathcal{Q}}_{nn,kk}^{\mathrm{TW}}(\mathbf{x}_0,\mathbf{p}_0;\mathbf{x}_t,\mathbf{p}_t),}
\end{equation}
where $\bar{C}_{nn,mm}^{\mathrm{TW}}(0)=1$.

For the $F=2$ case, the {TCF} with triangle window functions can be transformed to an equivalent form on a single $V_1(\mathbb{C}^2)$ phase space with any choice of $\gamma$ for pure electronic dynamics \cite{He2024}. The transformed {TCF} reads
\begin{equation}
    \begin{split}
    &{\Tr\left[\ketbra{n}{n} \hat{U}^\dagger(t)\ketbra{m}{m}\hat{U}(t)\right]}\\
    \mapsto&{\left(\bar{C}_{nn,mm}(t)\right)^{-1}\int_{\mathcal{S}(\mathbf{x}_0,\mathbf{p}_0;\gamma)}\dd\mu(\mathbf{x}_0,\mathbf{p}_0) \bar{\mathcal{Q}}_{nn,mm}(\mathbf{x}_0,\mathbf{p}_0;\mathbf{x}_t,\mathbf{p}_t)} \\
    \end{split}
\end{equation}
where
\begin{equation}
    \begin{split}
    &{\bar{\mathcal{Q}}_{nn,mm}(\mathbf{x}_0,\mathbf{p}_0;\mathbf{x}_t,\mathbf{p}_t)}\\
    =&{\left[2-\frac{(1+F\gamma)^2}{\min \{x_{n,0}^2+p_{n,0}^2,x_{m,t}^2+p_{m,t}^2\}^2 }\right]} \\
    &\times h\left(\frac{x_{n,0}^2+p_{n,0}^2}{2}-\frac{1+F\gamma}{2}\right) h\left(\frac{x_{m,t}^2+p_{m,t}^2}{2}-\frac{1+F\gamma}{2}\right) \\
    \end{split}
\end{equation}
and the time dependent normalization factor $\bar{C}_{nn,mm}(t)$ reads
\begin{equation}
    {\bar{C}_{nn,mm}(t) = \sum_{k=1}^2 \int_{\mathcal{S}(\mathbf{x}_0,\mathbf{p}_0;\gamma)}\dd\mu(\mathbf{x}_0,\mathbf{p}_0) \bar{\mathcal{Q}}_{nn,kk}(\mathbf{x}_0,\mathbf{p}_0;\mathbf{x}_t,\mathbf{p}_t).}
\end{equation}
This approach is a special case of a broader framework for the exact {population-population correlation function} for {the} two-state {quantum} system in ref. \cite{Cheng2024}.

We also propose a new type of positive semi-definite population-population correlation functions for general $F$-state systems, by introducing the \textbf{hill window function} {(HWF)}. The expression for $\bar{\mathcal{Q}}_{nn,mm}^{\mathrm{HW}}(\mathbf{x}_0,\mathbf{p}_0;\mathbf{x}_t,\mathbf{p}_t)$ reads
\begin{equation}
{\bar{\mathcal{Q}}_{nn,mm}^{\mathrm{HW}}(\mathbf{x}_0,\mathbf{p}_0;\mathbf{x}_t,\mathbf{p}_t)=\left[\hat{K}_\rho^{\mathrm{HW}}(\mathbf{x}_0,\mathbf{p}_0)\right]_{nn}\left[\hat{K}_A^{\mathrm{HW}}(\mathbf{x}_t,\mathbf{p}_t)\right]_{mm}},\label{eq:CF-HW}
\end{equation}
where the {HWF} reads
\begin{equation}
    {[\hat{K}_A^{\mathrm{HW}}(\mathbf{x},\mathbf{p})]_{mm}=\prod_{k\neq m}^F \left(\frac{x_m^2+p_m^2}{2}-\frac{x_k^2+p_k^2}{2}\right)^{B(F)}h\left(\frac{x_m^2+p_m^2}{2} - \frac{x_k^2+p_k^2}{2}\right),}
\end{equation}
with
\begin{equation}
    {B(F) = \frac{3}{7(F-1)} + \frac{60}{7(F+13)},}
\end{equation}
and
\begin{equation}
{[\hat{K}_\rho^{\mathrm{HW}}(\mathbf{x},\mathbf{p})]_{nn}=\prod_{k\neq n}^F h\left(\frac{x_n^2+p_n^2}{2} - \frac{x_k^2+p_k^2}{2}\right).}
\end{equation}
{Also, a time-dependent normalization factor}
\begin{equation}
    {\bar{C}_{nn,mm}^{\mathrm{HW}}(t) = \sum_{k=1}^F \int_{\mathcal{S}(\mathbf{x}_0,\mathbf{p}_0;\gamma)}\dd\mu(\mathbf{x}_0,\mathbf{p}_0) \bar{\mathcal{Q}}_{nn,kk}^{\mathrm{HW}}(\mathbf{x}_0,\mathbf{p}_0;\mathbf{x}_t,\mathbf{p}_t)}
\end{equation}
{is used.}

Appendix D shows that for the $F=2$ case, the above approach with the HWF belongs to a new theoretical framework for exact population-population correlation function in the frozen nuclei limit.  When only the $F=2$ case is considered, the HWF approach still involves a \textit{time-dependent} normalization factor for dynamics of composite/nonadiabatic systems, though it can be related to the work of ref. \cite{Mannouch2023} where a \textit{constant} normalization factor is used. The latter also satisfies the frozen nuclei limit \cite{Mannouch2023} for the $F=2$ case. Appendix D also offers a new theoretical framework for phase space approaches with non-covariant TCFs for general $F \ge 3$ case.

\section{Conclusion}\label{sec:final}

In this paper, we generalize the constraint phase space (CPS) formulation of refs. \cite{Liu2016,He2019,He2021a,He2022, Wu2024, Cheng2024, He2024} for discrete finite $F$-state quantum systems. Fully determined by the structure of the mapping kernel, the phase space is composed of several connected components, each of which has a geometric structure of a complex Stiefel manifold $\mathrm{U}(F)/\mathrm{U}(F-r)$ labeled by the maximal degeneracy degree of the eigenvalues of the corresponding mapping kernel. With no trajectory traveling from one connected component to another, different connected components are dynamically independent. By embedding the complex Stiefel manifolds into the Euclidean space, the generalized CPS can be described and parameterized.

The evolution of a trajectory on the generalized CPS is determined by the unitary transformation of the mapping kernel by the time evolution operator, which is also compatible with a symplectic structure on the Euclidean space containing the generalized CPS. By setting either the density kernel $\hat{K}_\rho(\mathbf{X})$ or the observable kernel $\hat{K}_A(\mathbf{X})$ to be covariant, the frozen nuclei limit is naturally satisfied when the exact mapping relation is satisfied at {time $t=0$}. There exist {various choices} of the kernels in the expression of the {TCF}. Typical types of correlation functions contain {the covariant-covariant (cc) type, covariant-noncovariant (cx) type, and noncovariant-covariant (xc) type.} {In addition to these} {TCF}s with covariant kernels, we also discuss and propose new expressions for window-window type population-population correlation functions. For each trajectory, the contribution to the phase space integral of the window-window type correlation function is positive semi-definite. This eliminates the existence of the negative non-physical population expectation values for composite systems. Recent works \cite{Wu2024, He2024} have demonstrated that both the EOMs and the integral expressions of {TCF}s are crucial for the accuracy of trajectory-based phase space dynamics methods, as also shown in the short-time analyses in refs. \cite{Shang2022thesis,Lang2024,Lang2022}. We believe that the generalized CPS framework will lead to useful expressions of {TCF}s not only for nonadiabatic transition systems but also quantum many-body systems { for spins/bosons/fermions}.

\section*{Declaration of competing interest}
There are no conflicts of interest to declare.

\section*{Acknowledgement}

We thank Baihua Wu, Xin He, Bingqi Li and Haocheng Lu for useful discussions. This work was supported by the National Science Fund for Distinguished Young Scholars Grant No. 22225304.

 \bibliographystyle{elsarticle-num} 
 \bibliography{references}





\end{document}